\documentclass[twocolumn]{pasj01}
\usepackage{stfloats}
\usepackage{natbib}

\begin{document} 
\Received{2019/04/17}
\Accepted{2019/06/06}
%\Published{yyyy/mm/dd}

\title{NH$_3$ Observations of the S235 Star Forming Region: Dense Gas in Inter-core Bridges}

%%% begin:list of authors
% Do NOT capitalize all letters in "textsc".
\author{Ross~A. \textsc{Burns}\altaffilmark{1,2,3,4}%
\thanks{Example: Present Address is xxxxxxxxxx}}
\email{ross.burns@nao.ac.jp}
\altaffiltext{1}{\small Mizusawa VLBI Observatory, National Astronomical Observatory of Japan, \\ 2-21-1 Osawa, Mitaka, Tokyo 181-8588, Japan}
\altaffiltext{2}{Korea Astronomy and Space Science Institute, 776 Daedeokdae-ro, Yuseong-gu, Daejeon, 34055, Republic of Korea}
\altaffiltext{3}{Joint Institute for VLBI ERIC, Oude Hoogeveensedijk 4, 7991 PD Dwingeloo, The Netherlands}
\altaffiltext{4}{Graduate School of Science and Engineering, Kagoshima University, 1-21-35 K\^orimoto, Kagoshima, Kagoshima 890-0065, Japan}
\altaffiltext{5}{Amanogawa Galaxy Astronomy Research Center, Kagoshima University, 1-21-35 K\^orimoto, Kagoshima, Kagoshima 890-0065, Japan}
\altaffiltext{6}{Ural Federal University, 19 Mira St. 620002, Ekaterinburg, Russia}
\altaffiltext{7}{Institute of Astronomy, Russian Academy of Sciences, 48 Pyatnitskaya Str. 119017, Moscow, Russia}
\altaffiltext{8}{Moscow Institute of Physics and Technology, 141701, 9 Institutskiy per., Dolgoprudny, Moscow Region, Russia}
\altaffiltext{9}{Mizusawa VLBI observatory, NAOJ 2-12, Hoshigaoka, Mizusawa, Oshu, Iwate 023-0861, Japan}
\altaffiltext{10}{Space Research Unit, Physics Department, North-West University, Potchefstroom, 2520, South Africa}
\altaffiltext{11}{Department of Physics and Astronomy, University of Nigeria, Carver Building, 1 University Road, Nsukka, 410001, Nigeria}
\altaffiltext{12}{Department of Physics, Nagoya University, Furo-cho, Chikusa-ku, Nagoya, Aichi 464-8601, Japan}
\altaffiltext{13}{Faculty of Science and Technology, Oita University, Oita 870-1192, Japan}

\author{Toshihiro \textsc{Handa},\altaffilmark{5}}
\email{Handa@sci.kagoshima-u.ac.jp }

\author{Toshihiro \textsc{Omodaka}\altaffilmark{4}}
\email{Omodaka@sci.kagoshima-u.ac.jp }

\author{Andrej~M. \textsc{Sobolev},\altaffilmark{6}}

\author{Maria~S. \textsc{Kirsanova},\altaffilmark{6,7,8}}

\author{Takumi \textsc{Nagayama},\altaffilmark{9}}

\author{James~O. \textsc{Chibueze},\altaffilmark{10,11}}

\author{Mikito \textsc{Kohno},\altaffilmark{12}}

\author{Makoto \textsc{Nakano},\altaffilmark{13}}

\author{Kazuyoshi \textsc{Sunada},\altaffilmark{9}}

\author{Dmitry~A. \textsc{Ladeyschikov},\altaffilmark{6}}

%%% end:list of authors

%% `\KeyWords{}' always has to be placed before `\maketitle'.
 %Do NOT move this preamble from here!

\maketitle

\normalsize

\begin{abstract}
Star formation is thought to be driven by two groups of mechanisms; spontaneous collapse and triggered collapse. Triggered star formation mechanisms further diverge into  cloud-cloud collision (CCC), ``collect and collapse" (C\&C) and shock induced collapse of pre-existing, gravitationally stable cores, or `radiation driven implosion' (RDI).
To evaluate the contributions of these mechanisms and establish whether these processes can occur together within the same star forming region we performed mapping observations of radio frequency ammonia, and water maser emission lines in the S235 massive star forming region. Via spectral analyses of main, hyperfine and multi-transitional ammonia lines we explored the distribution of temperature and column density in the dense gas in the S235 and S235AB star forming region.
The most remarkable result of the mapping observations is the discovery of high density gas in inter-core bridges which physically link dense molecular cores that house young proto-stellar clusters. The presence of dense gas implies the potential for future star formation within the system of cores and gas bridges.
Cluster formation implies collapse and the continuous physical links, also seen in re-imaged archival CS and $^{13}$CO maps, suggests a common origin to the molecular cores housing these clusters, i.e the structure condensed from a single, larger parent cloud, brought about by the influence of a local expanding H${\rm II}$ region. An ammonia absorption feature co-locating with the center of the extended H${\rm II}$ region may be attributed to an older gas component left over from the period prior to formation of the H${\rm II}$ region. Our observations also detail known and new sites of water maser emission, highlighting regions of active ongoing star formation.
\end{abstract}

\section{Introduction}
 
 %The understanding of star formation can be approached from Galactic scales such as the empirical relations of \cite{Schmidt59,Kennicutt98}, down to the scales of individual young stellar objects. Consequently, there must exist some causality acting upon star formation at all scales; from individual (proto-)stars ($< 10^1$ pc), local SFR-scales ($10^{1-2}$ pc), and Galactic-scale star formation ($> 10^2$ pc); see, for example \citet{Nguyen16}. The intermediate, SFR-scale investigations are therefore required to explore the influence of large scale phenomena on the small scale, and vice versa.

%While star formation on Galactic scales follows well behaved empirical relations \citep{Schmidt59,Kennicutt98}, and while star formation of individual young stellar objects (YSOs) is also well-studied, there have been fewer investigations linking molecular gas distributions with star formation at the intermediate scales, such as those of individual star forming regions (SFRs) within giant molecular clouds (GMCs). For Galactic-scale star formation relations such as the aforementioned Kennicut-Schmidt law to exist there must be some relation between star formation activities at all scales; from individual (proto-)stars ($< 10^1$ pc), local SFR-scales ($10^{1-2}$ pc), and Galactic-scale star formation ($> 10^2$ pc); see, for example \citet{Nguyen16}. The intermediate, SFR-scale investigations are therefore required to link our knowledge of star formation across the large- and small scales.

The existence of Galactic-scale star formation relations such as the Kennicut-Schmidt law \citep{Schmidt59,Kennicutt98} implies continuity in star formation activity at all scales; from individual (proto-)stars ($< 10^1$ pc), local star forming region (SFR)-scale ($10^{1-2}$ pc), and Galactic-scale star formation ($> 10^2$ pc); see, for example \citet{Nguyen16}. At an intermediate between the Galactic and (proto-)star scales, investigations on the scale of individual SFRs are therefore required for completeness; to link our knowledge of star formation across all scales.

%At these several-parsec scales there should exist an overlap between the relations found for the larger and smaller scales. This overlap should link the extended ambient gas density in a molecular cloud and the rate of forming YSOs. Furthermore, the star formation activity at these scales is affected by local phenomena such as mechanical and radiative feedback from massive stars, expanding supernova shells, in addition to the effects of the distribution and clumpiness of the medium from which the star forming cores condense gravitationally. 

At SFR scales, star formation is thought to be driven by two groups of mechanisms; spontaneous collapse and triggered collapse \citep{Elmegreen77,Elmegreen98}. Spontaneous collapse being the case when a super-critical density gas cloud is allowed to contract gravitationally, relatively undisturbed, while triggered collapse pertains to an external influence which encourages collapse by compressing sub-critical gas beyond critical density. Triggered star formation mechanisms further diverge into ``collect and collapse" (C\&C) \citep{Elmegreen77}, shock induced collapse of pre-existing, gravitationally stable cores, or `radiation driven implosion' (RDI) \citep{Sandford82} and large scale cloud-cloud collisions (CCCs) capable of generating the widespread over-densities required to drive sequential star formation \citep{Habe92,Haworth15a,Haworth15b,Torii17,Haworth18}.

The Ammonia molecule has long been recognised as a sensitive thermometer and densitiometer for probing the interstellar medium \citep{Ho83,Walmsley94}. It has been used extensively to probe the physical conditions in various stages of star formation including pre-stellar cores \citep{Ruoskanen11}, active star formation cores \citep{Harji91,Kir14,Kir16}, filementary structures \citep{Wu18} and large scale star formation surveys \citep{Friesen17}.

Several recent works by our group have used ammonia mapping observations to look for evidence of spontaneous and triggered star formation \citep{Toujima11,Chibueze13,Nakano17}, with the goal of uncovering which mechanism is dominant and whether these processes can occur together within the same star forming region.
As a continuation of this project, we conducted radio frequency ammonia transition mapping observations of the S235 ``main'' and S235AB (collectively termed `S235' hereafter) star forming region with the goal of mapping the physical conditions of molecular gas.

\begin{figure*}[ht]
\begin{center}
\includegraphics[scale=1.35]{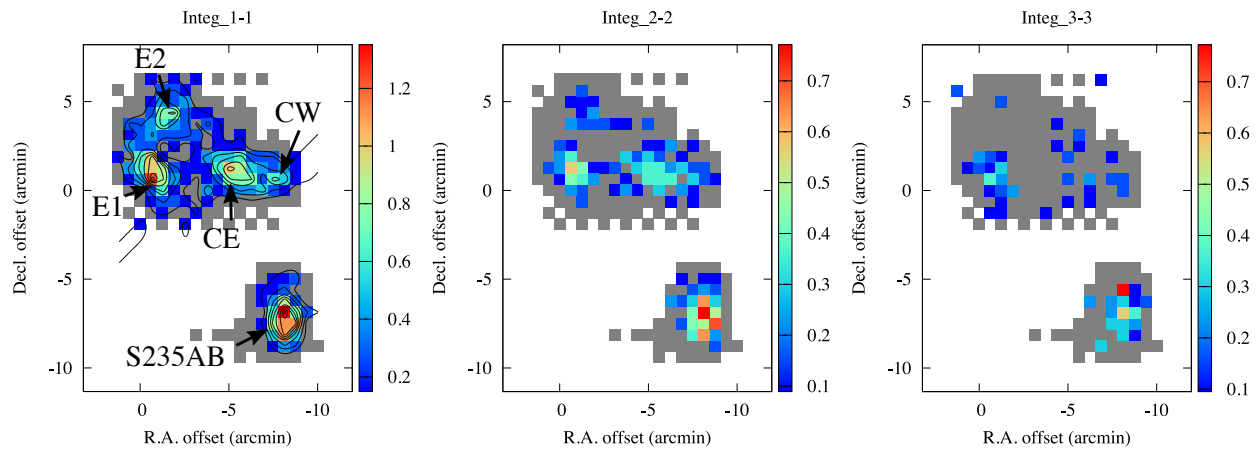}
\end{center}
\caption{Maps of the ammonia in S235 and S235AB, showing (from left to right) the intensities of the (1,1), (2,2) and (3,3) emission. Colour scales indicate brightness temperatures in units of Kelvin, scaled individually for each map. Grey squares indicate no emission at or above the 2$\sigma$ cutoff. Contours in the (1,1) map increase from 3 times the rms noise in integer intervals.
\label{112233}}
\end{figure*}

S235 is the most active region of star formation of the G174+2.5 giant molecular cloud. 
It houses multiple dense gas cores which have been extensively studied using NH$_3$, CS and $^{13}$CO molecular lines \citep{Kir08,Kir14}. However, previous NH$_3$ maps of S235 were exclusive to the well-known dense cores and did not sample the regions between and around the cores. More complete observations were made by \citet{DewanganOjha17} using CO and its isotopologues, thus tracing he widespread diffuse gas.
% and were therefore not sampling inter-core gas which is free from the influence of internal star formation feedback. 
Regarding continuum emission, S235 is home to a circular H${\rm II}$ region driven by ionizing radiation from an O9.5 V star, (BD +35$^{\circ}$1201) \citep{Georgelin73}, at the center of the H${\rm II}$ region. \citet{Dew11} report that stellar densities of YSOs in S235 concentrate in the four molecular cores which are referred to as East1, East2, Central-West and Central-East in \citet{Kir08,Kir14}. In this work we adopt their notation and use short-hands E1, E2, CW and CE, respectively. \citet{Kir08,Kir14} suggest that the molecular cores may have formed via the C\&C mechanism through interaction with the H${\rm II}$ region. 
More recently, work by \citet{Dew16} have added support to this picture by showing with high confidence that star formation in S235 is driven by interaction with the expanding H${\rm II}$, as is the conclusion of their thorough multi-wavelength investigation.
\citet{DewanganOjha17} further revealed evidence of a past CCC event which likely initiated the subsequent star formation observed today.

Further to the south, the S235AB region is home to a younger H${\rm II}$ region, S235A, and very intense star formation indicated by high concentrations of YSOs \citep{Dew11}. Maser activity in this region also points to the presence of very young massive star formation \citep{Felli07,Burns15a}, and enabled the distance to S235 to be established as $D_\pi = 1.56 ^{+0.09} _{-0.08}$ kpc, via maser parallax \citep{Burns15a}. The combined presence of molecular cores, H${\rm II}$ regions and star formation at various evolutionary stages makes S235 an ideal region to investigate various scenarios pertaining to spontaneous and triggered star formation.

\section{Observations and archival data}

Observations of S235 were carried out between December 2013 and June 2014 using the Nobeyama 45-m radio telescope, operated by the Nobeyama Radio Observatory (NRO), a branch of the National Astronomical Observatory of Japan (NAOJ). 
The telescope was operated in single side-band mode with frequency windows centered at the rest frequencies of NH$_{3}$ inversion transitions (J,K) = (1,1), (2,2) and (3,3) at 22.6914, 23.722 and 23.870 GHz, respectively, and the H$_{2}$O $6_{12} - 5_{23}$ maser transition at 22.235 GHz. All frequency bands were observed simultaneously with dual linear polarisations and autocorrelated with a 0.38 km s$^{-1}$ velocity channel spacing. The FWHM beamsize was 75" and pointing was checked every 1 $\sim$ 2 hours and deviations were kept below 5".

Mapping observations were conducted in position switching mode (ON-OFF) using map grid spacings of 37.5". Sky subtraction was achieved by observing a region with no emission. Repeated integrations of 20 seconds were made at each point, with 3 ON points for every OFF point.
$T_\mathrm{sys}$ varied between $T_\mathrm{sys} = 90 \sim 140$ K for all observing runs, thus scans were integrated until an \textit{rms} noise level of 0.04 K was reached for each mosaic point. This provided an overall consistency in map noise irrespective of the changing $T_\mathrm{sys}$.
Maps were created sequentially, beginning at the coordinates of the dense gas cores described in the literature (E1, E2, Central, AB) and extending outwards from those cores until no emission is detected. It is possible that ammonia emission existing between the S235 complex and S235AB was missed. Confirmation should be made by further observations.
The total observing time required to produce the final maps of the S235 and S235AB regions was 120 hours.

%S235, 181 x 30min = 90hrs

%S235AB, 58 * 30min = 30 hrs

Data reduction was performed using the NEWSTAR software, which has been developed and maintained by NRO. Baseline subtraction was performed individually for all scans, frequency bands and polarisations, after which polarisations were combined to Stokes $\rm I$. 
Fitting and analyses of the molecular inversion spectrum of ammonia were carried out with \textit{gnuplot} routines which paramaterised the main line and satellite line profiles. Non-detections (grey squares in the emission maps) were recorded lacking a 2$\sigma$ detection.

%To supplement our data maps we revisit the CS$(2-1)$ and $^{13}$CO$(1-0)$ data published by \citet{Kir08}. These data were taken with the 20-m Onsala Space Observatory with a FWHM beam of 40$^{\prime \prime}$ with single grid spacing - resulting in a resolution comparable to our own observations.

\section{Results}
\label{RES}
\subsection{Ammonia: Emission maps}
Maps of the ammonia (1,1), (2,2) and (3,3) emission in S235 are shown in Figure~\ref{112233} where the reference coordinate (0,0) corresponds to 
$(\alpha, \delta)_{\mathrm{J}2000.0}=(05^{\mathrm{h}}41^{\mathrm{m}}33^{\mathrm{s}}.8$, +35$^{\circ}$48'27"). Ammonia from the (1,1) transition was detected in our observations at all cores, the regions between the cores, and in S235AB. Emission from the (2,2) transition was seen in the 4 cores and S235AB but was rarely detected in the inter-core regions. Emission from the (3,3) transition was primarily detected in East 1 and S235AB.

\begin{figure*}[!ht]
\begin{center}
\includegraphics[scale=1.0]{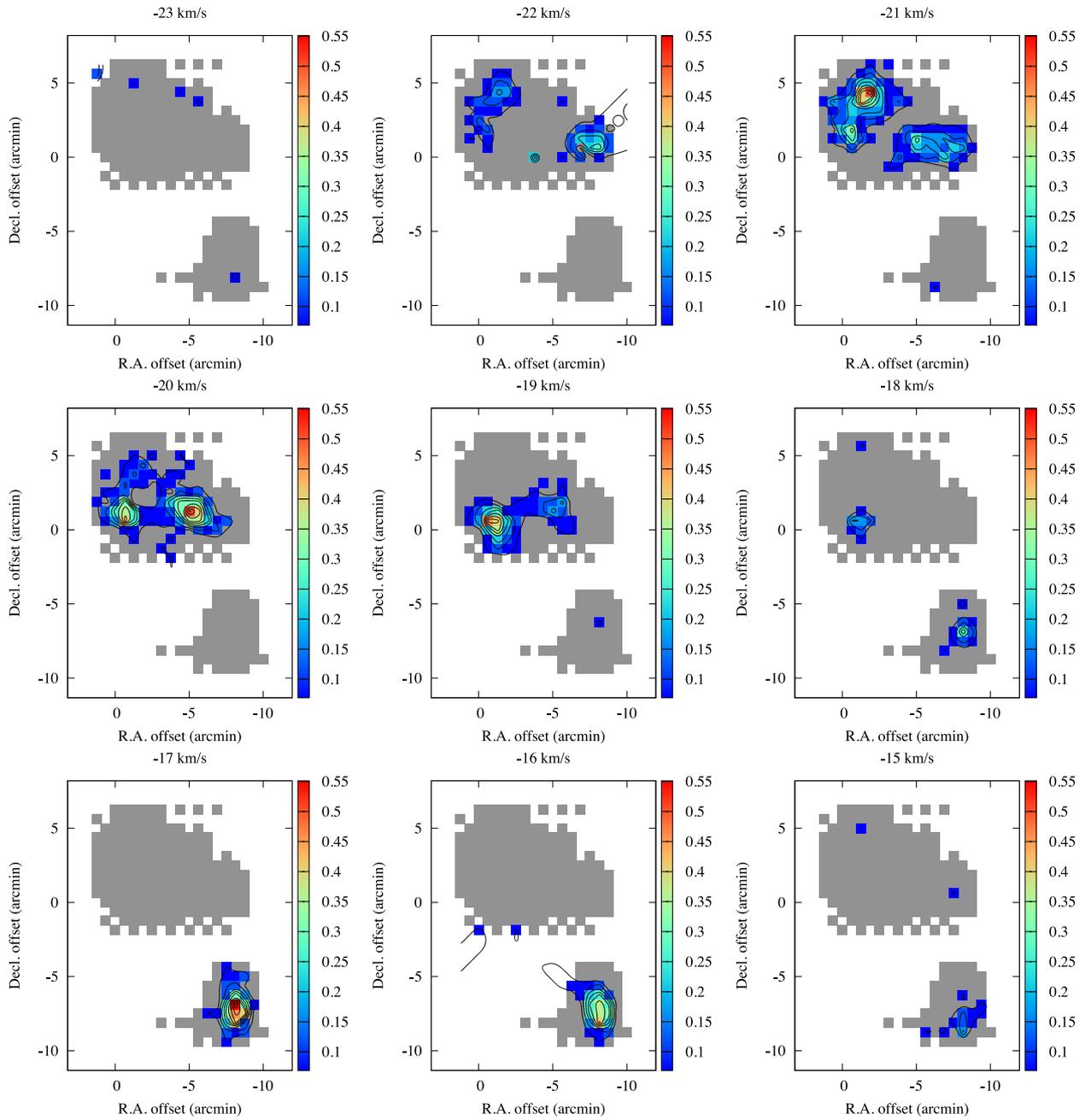}
\end{center}
\caption{Channel maps of the ammonia (1,1) emission in S235 and S235AB. Colour scales indicate brightness temperatures in units of Kelvin, scaled individually for each map. Contours in the (1,1) map increase from 3 times the rms noise in integer intervals. Grey squares indicate no emission at or above the 2$\sigma$ cutoff.
\label{VEL}}
\end{figure*}

\begin{figure*}[!ht]
\begin{center}
\includegraphics[width=0.79\textwidth]{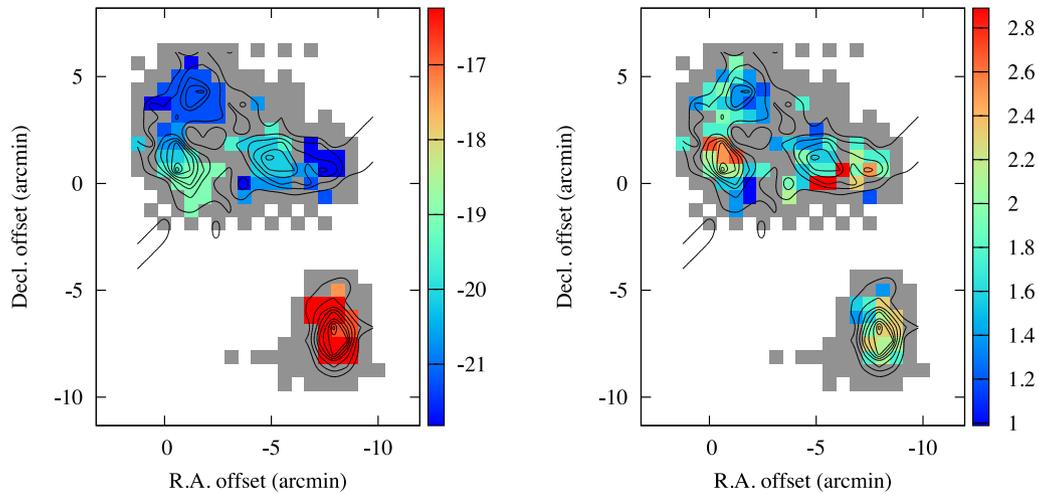}
\end{center}
\caption{The first and second moment maps of ammonia in S235 and S235AB, shown \emph{left} and \emph{right}, respectively. Colour scales are in units of km s$^{-1}$. Overlain contours outline the ammonia (1,1) intensity profile from Figure~\ref{112233}
\label{VVWC}}
\end{figure*}

Channel maps of integer increment are shown in Figure~\ref{VEL}. Core and inter-core gasses are continuous in velocity with emission velocities ranging from -15 to -21 km s$^{-1}$ (Figure~\ref{VVWC}). Thus the majority of emission is seen to be blueshifted with respect to the ambient molecular cloud at -17 km s$^{-1}$ \citep{Heyer96}. Velocity widths of the ammonia spectra are in the range of 1 to 3 km s$^{-1}$.

\begin{figure*}[!hb]
\begin{center}
\hspace{+0.4cm}
\includegraphics[width=0.82\textwidth]{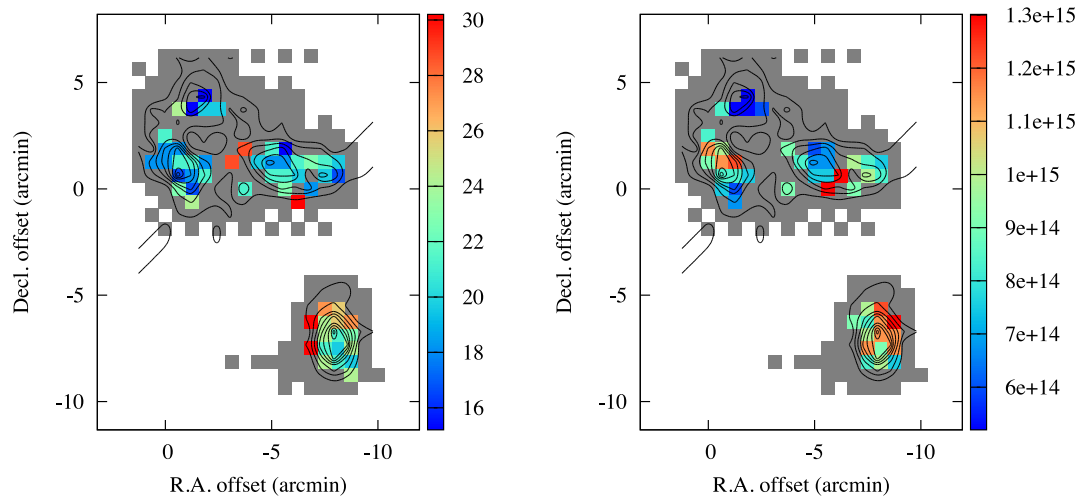}
\end{center}
\caption{Maps of the physical gas parameters in S235 and S235AB. \emph{Left} shows the rotational temperature, $T_{rot}$, of ammonia in which the colour scale indicates temperatures in Kelvin, and, (\emph{Right}) shows the total column density, $N_{TOT}$, of ammonia gas in units of cm$^{-2}$.
\label{PhysPrms}}
\end{figure*}

\subsection{Deriving physical parameters of molecular gas}
\label{DER}
To derive the physical conditions of the ammonia gas in our mapped regions we first calculate the optical depth, $\tau$, using the ratio of integrated main and satellite line temperatures:

\vspace{-0.3cm}

\begin{eqnarray}
\frac{T_{MB}(main)}{T_{MB}(satellite)} = \frac{1 - e^{-\tau}}{1 - e^{-a\tau}}
\end{eqnarray}

\noindent Where $a$ is the natural intensity ratio of the satellite to main lines. We then calculate the rotation temperature of the gas, $T_{rot}$, using:

\vspace{-0.3cm}

\begin{eqnarray}
{\scriptstyle T_\mathrm{rot} = -41.5 ~/~ \ln \left( \frac{-0.282}{\tau} \times ln \left[ 1- \frac{T_{MB}(2,2)}{T_{MB}(1,1)} \times ( 1 - e^{-\tau} ) \right] \right) }
\end{eqnarray}

\noindent The column density of gas emitting at the (J,K) transition, $N(J,K)$, is calculated by:

\vspace{-0.3cm}

\begin{eqnarray}
N(1,1) = A \times 10^{13}~ \tau(J,K,main)   ~T_{rot} ~\Delta \mathrm{v}_{1/2}
\end{eqnarray}

\noindent Where $a$ in this case is $A$ dimensionless constant corresponding to $A = 2.78, 1.31, 1.03$ for $(J,K)=(1,1), (2,2) , (3,3)$, respectively. 
Finally, the total NH$_3$ column density, $N_{TOT}$, is calculated assuming local thermodynamic equilibrium, using:

\begin{eqnarray}
{\scriptstyle N_\mathrm{TOT} = \frac{N(J,K)}{g_J, g_I, g_K}~ \exp \left[\frac{E(J,K)}{T_{rot}}\right] \times \Sigma ~g_J, g_I, g_K~ \exp \left[\frac{-E_{i}(J,K)}{T_{rot}}\right] }
\end{eqnarray}

Where $g_J$ is the rotational degeneracy, $g_I$ is the nuclear spin degeneracy, and $g_K$ is the K-degeneracy.

\small
\begin{table*}[tp]
\caption{Physical conditions of the star forming cores in S235.\label{PHYS}}
\begin{center}
\begin{tabular}{cccccc}
\hline
Name		&$\tau$ &$T_{rot}$ & $\Delta \mathrm{v}$ &$ N_{H_2}$& $M_{LTE}$\\
& 	    &  [K]   &  [km s$^{-1}$] & [cm$^{-2}$]& $M_{\odot}$\\
\hline
East 1    & $0.28 \pm 0.32$ & $19 \pm 1$ & $1.75 \pm 0.47$  & $(8.7 \pm 1.8) \times 10^{21}$ & $138\pm11$\\ 
East 2    & $0.41 \pm 0.53$ & $18 \pm 2$ & $1.56 \pm 0.23$  & $(10.6 \pm 1.5) \times 10^{21}$ & $55\pm6$\\ 
Central E & $0.62 \pm 0.46$ & $20 \pm 2$ & $1.77 \pm 0.41$  & $(1.9 \pm 0.4) \times 10^{21}$ & $152\pm12$\\ 
Central W & $1.43 \pm 0.56$ & $20 \pm 1$ & $2.08 \pm 0.37$  & $(5.4 \pm 0.7) \times 10^{21}$ & $101\pm9$\\ 
S235AB    & $0.08 \pm 0.23$ & $23 \pm 1$ & $1.94 \pm 0.31$  & $(3.2 \pm 0.4) \times 10^{21}$ & $276\pm26$\\ 
\hline
\hline
\end{tabular}
\end{center}
\end{table*}
\normalsize

\begin{figure*}[!hb]
\begin{center}
\includegraphics[angle=270,width=0.57\textwidth]{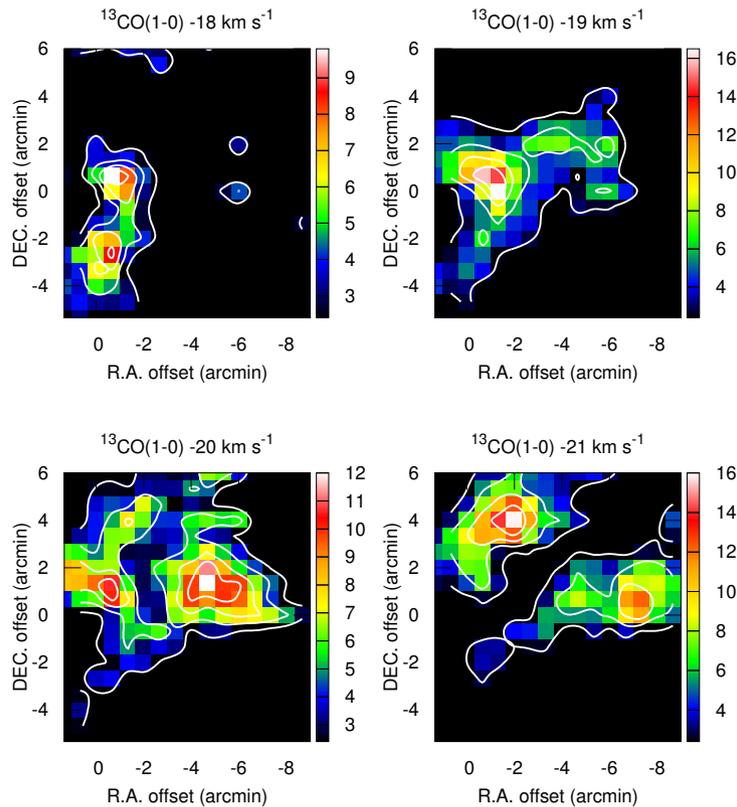}
\end{center}
\caption{Re-imaged $^{13}$CO data of S235 (excluding S235AB) from \citet{Kir08} where colour indicates brightness temperature, $T_\mathrm{b}$. Coordinate offsets match those of the Ammonia maps\label{13CO}}
\end{figure*}

\subsection{Ammonia: Mapping the physical conditions of molecular gas in S235}

Based on the relations described in Section~\ref{DER} we estimated the physical conditions of ammonia gas at each point in the grid. Since only a few map positions had sufficiently bright satellite emission for deriving the gas opacity we adopt the average value of $\tau$ for each core for the subsequent calculations of $T_{rot}$ and $N_{TOT}$ at each map point in S235 and S235AB. Maps of these parameters are shown in Figure~\ref{PhysPrms}.

We then evaluated further physical parameters (gas opacity, $\tau$; rotational temperature, $T_{rot}$; velocity width, $\Delta \mathrm{v}$; Total gas column density, $N_{H_2}$; and core mass, $M_{LTE}$) associated with cores E1, E2, CW, CE and S235AB. For this step we employed an ammonia abundance ratio of $X(\rm NH_3) = 1.379 \times 10^{-7}$, following \citet{Millar97}. Physical parameters for cores were reached by integrating over their angular area, based on their sizes given in \citet{Kir14}. These are reported in Table~\ref{PHYS}.

\subsection{CS and $^{13}$CO: Re-imaging archival maps}

\begin{figure*}[!ht]
\begin{center}
\includegraphics[angle=270, width=0.57\textwidth]{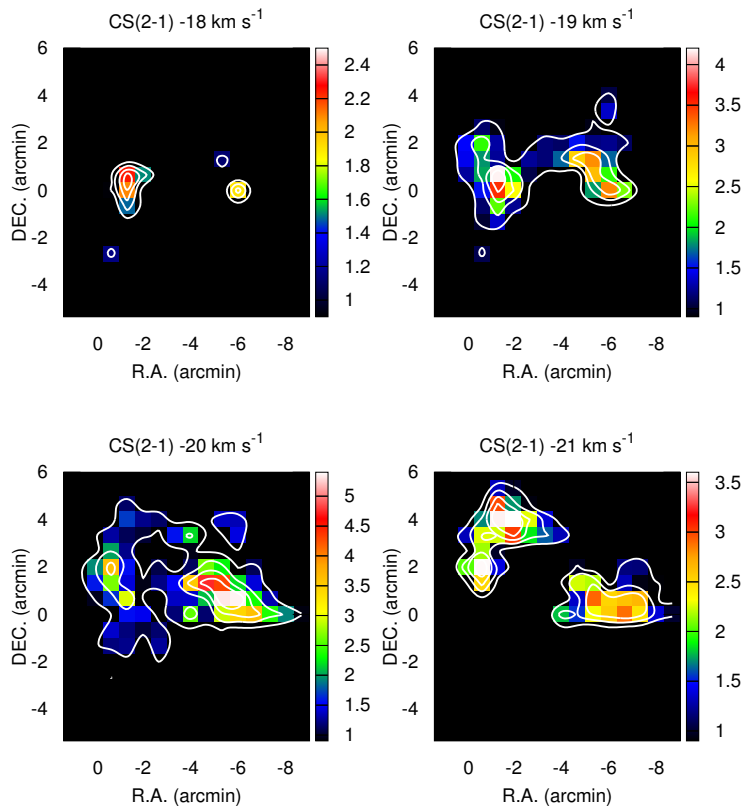}
\end{center}
\caption{Re-imaged CS data of S235 (excluding S235AB) from \citet{Kir08} where colour indicates brightness temperature, $T_\mathrm{b}$. Coordinate offsets match those of the Ammonia maps\label{CS}}
\end{figure*}

In order to compare the results of our ammonia mapping observations with the distributions of gasses at other densities, we revisited the CS$(2-1)$ and $^{13}$CO$(1-0)$ data of \citet{Kir08}. The maps, made with the Onsala 20-m telescope, have FWHM beamsizes of 34$^{\prime \prime}$ and 38 $^{\prime \prime}$ for the CS and $^{13}$CO data, respectively, producing maps of almost equal angular resolution as the ammonia grid spacings taken at Nobeyama. The reader may refer to the aforementioned publication for further details of the observations. Channel maps were produced at integer velocity intervals, these are shown in Figures~\ref{13CO} and \ref{CS}.

\subsection{Inter-core gas bridges: physical parameters}
\label{SPECT}
From Figure~\ref{112233} it is apparent that an extended gas component of ammonia was detected in the regions outside of the main cores apparently forming a network of inter-core bridges. Withholding a deeper discussion of these gas bridges for Section~\ref{DISCUSSION}, in this section we derive the physical parameters of gas belonging to this component.

\begin{figure}[!hb]
\begin{center}
\hspace{-1cm}
\includegraphics[scale=0.7]{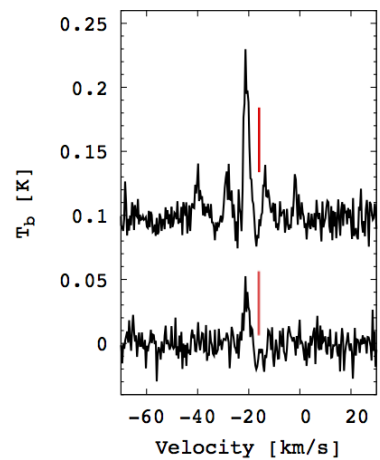}
\end{center}
\caption{Integrated spectra of inter-core ammonia gas, showing the (1,1) transition (\emph{above}) and the (2,2) transition (\emph{below}). \label{InterCore11}}
\end{figure}

Ammonia emission from the `inter-core' gas were too weak to analyse on a point-by-point basis. We therefore integrated signals from all pointings considered to be inter-core gas by the definition of being outside the derived core radii listed in Table~\ref{PHYS}. Since inter-core gas exhibits little to no star formation activity, no large velocity widths or gradients, such an integration can be considered reliable. Integrated spectra of the ammonia (1,1) and (2,2) from inter-core gas are shown in Figure~\ref{InterCore11}. 
%and ~\ref{InterCore22}, respectively.

Repeating the spectral analyses outlined in Section~\ref{RES} we derived the physical properties of the inter-core gas in S235. For this gas component we estimated an optical depth of $\tau = 0.12 \pm 0.43 $, rotation temperature of T$_{rot} = 18.01 \pm 3.78$ K, and a total gas column density of N$_{H_2} = 1.3 \pm 4.8 \times 10^{21} $ cm$^{-2}$. Assuming that the gas bridges are of equal depth as their width (0.4 pc) the density of gas in the gas bridges would be $\rho= 1.1 \pm 4.1 \times 10^{3}$ cm$^{-3}$.

Filamentary clouds become unstable when the line mass per unit length exceeds the critical line mass, i.e.  ${M}_{{\rm{line}}}>{M}_{{\rm{crit}}}$ (see \cite{Inutsuka97} and, for example, \cite{Ryabukhina18}). The parameters above imply a line mass of ${M}_{{\rm{line}}} \sim 85 M_{\odot}/pc$. The critical line mass depends only on gas temperature and mean molecular mass (eq. 60 of \cite{Ostriker64}) thus the 18 K gas bridges have ${M}_{{\rm{crit}}} \sim 35 M_{\odot}/pc$. Consequently, in the absence of support, the gas bridges would be gravitationally unstable.

%column density of N$_{[1,1]} = 5.58 \times 10^{13} \pm 20.32$ cm$^{-2}$, 

%By fitting the integrated ammonia spectrum in the same way as previously it was possible to derive the physical properties of the inter-core gas in S235. For this gas component we estimate an optical depth of $\tau = 0.119241 \pm 0.4337 $, rotation temperature of T$_{rot} = 18.009 \pm 3.78232$ K, column density of N$_{NH_3} = 5.58 \times 10^{13} \pm 20.32$ cm$^{-2}$, and a total gas column density of N$_{TOT}= 1.29 \times 10^{14} \pm 4.78$ cm$^{-2}$. Assuming that the gas bridges are of equal depth as their width the density of gas in the gas bridges would be $\rho= 1.33 \times 10^{-4} \pm 4.93$ cm$^{-3}$.

%estimate total density of bridges, assuming that they are as deep as they are wide. 
%Dividing the column density by the depth (assumed to equal the width; depth = 0.4pc [1,234e+18])
%=1.04 e-4 if it were a square pipe
%times by the ratio of crossection area of square/circle of equal radius
%= 1.33 e-4 cm-3

% \newpage

\subsection{A low velocity gas component seen in absorption}
\label{ABS}

\begin{figure*}[!ht]
\begin{center}
\includegraphics[width=0.79\textwidth]{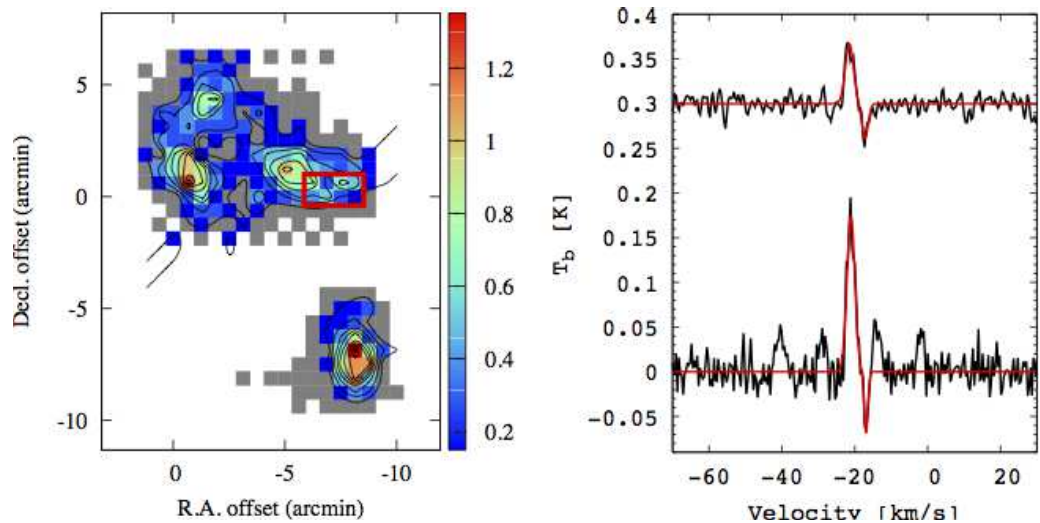}
\end{center}
\caption{\emph{Left}: The region of points included in the integrated spectrum analysis of the absorption signature, highlighted with a red rectangle. \emph{Right}: Integrated spectra of (\emph{below}) ammonia (1,1) and (\emph{above}) ammonia (2,2) gas spatially consistent with the extended HII region. Data were Hanning smoothed with a 3-point window. The red line shows Gaussian profiles fit to the emission and absorption peaks.\label{REGION}}
\end{figure*}

Another interesting feature in the ammonia data became apparent from Figure~\ref{InterCore11}; the integrated ammonia spectra produced from the combined inter-core pointings exhibits a tentative absorption signature at the velocity marked by the red line - prompting deeper investigation of ammonia gas near the H${\rm II}$ continuum peak. To clarify the case of the suspected absorption signature we integrated ammonia spectra from a $4 \times 2$ pointing region which is spatially consistent with the H${\rm II}$ region to increase the signal to noise ratio of the spectrum. The selected regions are indicated by the red rectangle locus in Figure~\ref{REGION}-\emph{left}. The resulting spectra are shown in Figure~\ref{REGION}-\emph{right} for the ammonia (1,1) and (2,2) gas.

%Selected pointings were restricted to the bright blueshifted ($V_{\rm LSR} < -20$ km s$^{-1}$) gas (see Fig~\ref{VEL}) to avoid overlapping in velocity with the absorption signature (around $V_{\rm LSR} \sim -16$ km s$^{-1}$).

The integrated spectra show a clear absorption signature in both the (1,1) and (2,2) gas, no absorption was seen in the (3,3) transition. To test whether this could be an artifact caused by the presence of weak ammonia emission in the reference OFF-point, we integrated spectra from the S235AB region which is not associated with an extended HII region. A contaminated OFF-point would effect all map points in our observations (both S235 and S235AB) equally, however no such absorption signature was found in the integrated S235AB spectrum, supporting the authenticity of the ammonia absorption near the S235 H${\rm II}$ region.

From the integrated spectrum we derive a brightness, peak velocity, and velocity half width at half maximum of $T_{b} = 0.17 \pm 0.01$ K, $\mathrm{v}_{LSR} = -21.07 \pm 0.08$ km s$^{-1}$ and $\Delta \mathrm{V}_{HWHM} = 1.29 \pm 0.10 $ km s$^{-1}$ in emission, and $T_{b} = 0.07 \pm 0.02$ K, $\mathrm{V}_{LSR} = -16.89 \pm 0.14$ km s$^{-1}$ and $\Delta \mathrm{V}_{HWHM} = 0.59 \pm 0.02 $ km s$^{-1}$ for the absorbing gas. The two gas components are sufficiently separated in velocity such that their spectral profiles do not interfere significantly.

Compared to individual points in the searched region, the absorption signature was \emph{enhanced} when multiple pointings were integrated, indicating that the absorbing gas component has an extent that is larger than a single beamsize. As such, we rule out interpretations that invoke compact sources of localised absorption of foreground gas on a background of line emission (such as a P-Cygni profile) instead, an interpretation involving two distinct ammonia gas components is preferred; one at $-17$ km s$^{-1}$, seen in absorption, and one at $-21$ km s$^{-1}$, seen in emission. A similar scenario with more clearly defined spectra is shown in \citep{Wilson78} who also reach an interpretation of multiple extended gas components. The production of either an emission or an absorption line signature must come from differences in the properties of the gas components themselves, i.e: $T_{line(-17km/s)} < T_{cont} < T_{line(-21km/s)}$. Here, T$_{cont}$ is the continuum brightness temperature and T$_{line}$ is the brightness temperature of molecular line emission whose velocity is indicated in subscript parentheses.

The brightness temperature of the continuum emission cannot be obtained directly from our data since all spectra required spectral baseline fitting during the data reduction stage. As such we must estimate $T_{cont}$ based on observations reported in the literature. 
First, we estimate the optical depth of the continuum emission at the frequency of our ammonia observations via $\tau = 3.28 \times 10^7 ~ (T_e)^{-1.35}  (\nu)^{-2.1} (EM)$. 
Where $T_e$ is the electron temperature, typically taken as 10,000 K for H${\rm II}$ regions, $\nu$ is set to the ammonia (1,1) transition frequency, 22.6914 GHz, and $EM$ is the emission measure.
The brightness temperature of the continuum emission can then be estimated using 
$T_{b} = T_{e} (1 - e^{-\tau})$.

\citet{Israel78} measure an emission measure of S235 to be $E M = 0.8 \times 10^4$ pc cm$^{-6}$ by observations at 1415 MHz with the Westerbork interferometer. On the other hand, \citet{Silverglate78} calculated $E M = 3.7 \times 10^4$ pc cm$^{-6}$ at 2371 MHz using the 305 m telescope at Arecibo. 
Employing values from Isreal et al. and Silverglate \& Terzian., respectively, we estimate a range of brightnesses of T$_{b} = 0.036$ and $0.17$ K for the continuum emission in S235; the intermediate brightness temperature of the continuum emission is capable of explaining the detection of both emission and absorption of ammonia in S235 for the (1,1) transition.

% The difference in these values may come from differences in the nature of interferometers and single-dish observations, i.e. the issue of lower flux recovery for interferometers observing extended sources. 

%s_e             = 1.08575          +/- 0.08151      (7.507%)
%s_a             = 0.503305         +/- 0.1393       (27.68%)
%vel_e           = -21.0654         +/- 0.08124      (0.3857%)
%vel_a           = -16.8924         +/- 0.1391       (0.8237%)
%amp_e           = 0.173717         +/- 0.01126      (6.481%)
%amp_a           = 0.0691233        +/- 0.01652      (23.91%)

%s_e             = 1.08688          +/- 0.1418       (13.05%)
%s_a             = 0.316689         +/- 0.0752       (23.75%)
%vel_e           = -21.1329         +/- 0.1417       (0.6707%)
%vel_a           = -16.9952         +/- 0.0792       (0.466%)
%amp_e           = 0.0689386        +/- 0.007785     (11.29%)
%amp_a           = 0.0673742        +/- 0.01416      (21.01%)

% WITH HANNING 3
%s_e             = 1.12973          +/- 0.09319      (8.249%)
%s_a             = 0.843834         +/- 0.1373       (16.27%)
%vel_e           = -21.4861         +/- 0.09031      (0.4203%)
%vel_a           = -17.1786         +/- 0.1345       (0.783%)
%amp_e           = 0.0667873        +/- 0.004605     (6.895%)
%amp_a           = 0.0389135        +/- 0.005304     (13.63%)

The brightness temperatures of the (2,2) transition lines were T$_{b} = 0.07 \pm 0.01$ K in emission and T$_{b} = 0.04 \pm 0.01$ K in absorption. Repeating the continuum brightness calculations at the frequency of the (2,2) emission, 23.722 GHz, gives a range of $T_\mathrm{b} = 0.034$ to $0.157$ K. Although the range of values of T$_{cont}$ is consistent with being able to explain the emission and absorption in the (2,2) emission, the weaker ammonia (2,2) emission makes the claim less certain.

Another, more direct conclusion drawn from the presence of an absorption spectrum is that the $-17$ km s$^{-1}$ component must be in the foreground. Since the present literature finds consensus regarding the blueshifted emission to be foreground gas, we can confirm the red shifted gas to be a second foreground gas component in S235.

\subsection{H$_2$O maser emission: Signposts of star formation}

\begin{figure}[!ht]
\begin{center}
\hspace{-1cm}
\includegraphics[scale=0.7]{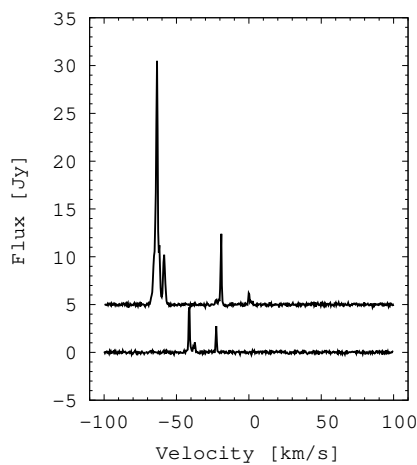}
\end{center}
\caption{H$_2$O maser emission detected in E1 (\emph{lower}) and S235AB (\emph{upper}). \label{MASERS}}
\end{figure}

In our observations, water maser emission was detected in E1 and S235AB. No previous records of water maser detections exist for E1, therefore these represent newly detected maser features. Emission spectra are shown in Figure~\ref{MASERS}, where a conversion of 2.7 Jy/K for the Nobeyama telescope has been applied.

The new maser at in E1 (Figure~\ref{MASERS}, \emph{lower}) was brightest at map grid $(\alpha, \delta)_{\mathrm{J}2000.0}=(05^{\mathrm{h}}41^{\mathrm{m}}31^{\mathrm{s}}.3$, +35$^{\circ}$50'19.5"). One 2.6 Jy emission peak was detected at a velocity of $-22.5$ km s$^{-1}$, which is consistent with the velocity of the E1 core ammonia gas. At least two more maser velocity components were found at -37.3 and -41.0 km s$^{-1}$ which are blueshifted with respect to the core gas, having fluxes of 1.0 and 4.7 Jy, respectively.
This maser is situated near the most luminous of embedded YSOs in E1 which were identified in the \emph{Spitzer} SEDs of \citet{Dew11}. Furthermore, the emission spectrum is that of a dominant blue-shift maser (DBSM). Such maser sources are thought to be associated with jets \citep{CP08,Motogi13,Motogi15,Burns15a}, indicating active star formation.

%4.5 and 0.8 Jy, respectively.

%Another new source of water maser emission was detected in Central-East (Figure~\ref{MASERS}) consisting of a single component with a flux of 0.5 Jy. Additional 2-channel binning was required to clarify the detection by increasing the signal to noise ratio. The maser velocity was [-17] km s$^{-1}$ which is slightly redshifted compared to the Central-East core gas, but constitutes a comparably low radial velocity for a water maser.

Maser emission was also detected in the S235AB region (Figure~\ref{MASERS}, \emph{upper}). The maser in S235AB is also a DBSM source with velocity components near 0, $-20$, and $-60$ to $-70$ km s$^{-1}$. These maser velocity components were previously catalogued as part of the Medicina patrol discussed in \citet{Felli07}. Water masers in S235AB are known to be associated with massive star formation in that region and are discussed in the context of multi-epoch VLBI observations in \citet{Burns15a}.

% \newpage

\section{Discussion}
\label{DISCUSSION}
%\subsection{The formation of the S235 star forming region} 
%In this section we discuss the possible origins of the inter-core gas bridges linking the star forming clumps in S235, and the corresponding implications on possible formation scenarii of the S235 star forming regions.

\subsection{Ammonia emission and absorption: young and old gas components in S235}

The absorption signature discussed in Section~\ref{ABS} suggests the presence of two velocity components of molecular gas; that seen in emission at -21 km s$^{-1}$ and that seen in absorption at -17 km s$^{-1}$. Since the differences in emission/absorption come from the physical properties of the gas we can infer that the -17 km s$^{-1}$ component has a lower excitation temperature than the -21 km s$^{-1}$ component. The large-scale molecular gas cloud in which S235 is embedded has a velocity of -16 km s$^{-1}$ \citep{Heyer96,Kir08}. In agreement with the conclusions of \citet{Kir14}, we speculate that the -17 km s$^{-1}$ absorbing gas component represents the remnants of the progenitor cloud, existing prior to- and yet uninfluenced by the formation of the H${\rm II}$ region - hence its low brightness temperature, opacity and density. On the other hand, the relatively enhanced -21 km s$^{-1}$ component, seen in emission, traces gas that is being heated by interaction with the H${\rm II}$ region.

Using $^{12}$CO and $^{13}$CO line data, \citet{DewanganOjha17} investigated the molecular boundaries of two clouds associated with the S235 and S235ABC regions, where two velocity components were traced.
The region of ammonia absorption co-locates with one intersect of the two CO clouds of \citet{DewanganOjha17} (see their Figures 5 and 6). They also reported active star formation toward these boundaries. The evidence suggests that the two gas components seen in ammonia may have been involved in a previous CCC event.

\subsection{Star formation activity traced by water maser emission}

Our observations detected two sites of water maser emission, one of which being a new detection. Water masers indicate that E1 and S235AB are active sites of star formation. This is not surprising since these cores have been discussed extensively in the context of their star formation activity in several previous publications \citep{Kir08, Kir14, Dew11, Dew16}. The maser detections are consistent with the view that both cores are relatively young members of the complex, and open the opportunity to perform further high resolution VLBI studies of the star formation activity in E1.

\subsection{Ammonia gas bridges as remnants of induced fragmentation}

The induction of recent star formation in S235 via the influence of the expanding H${\rm II}$ region has been explored and supported by several previous works \citep{Kir08,Dew16,Bieging16}. To supplement these previous works without repeating them we concentrate our discussion on the inter-core gas discovered in our observations to discuss what appear to be gas remnants of triggered fragmentation. 

Inter-core gas bridges were reported in \citet{Salii02}, and were first seen in the S235 star forming region in \citet{DewanganOjha17} who report a broad bridge feature of CO gas. Such features (seen in PV) can be produced via the CCC process REF. \citet{DewanganOjha17} also discuss CCC as a possible formation scenario in the S235 and S235ABC complex.

While optically thin CO gas traces regions of high column density, ammonia emission has a higher critical density and thus traces high density gas.
Our observations reveal the presence of dense molecular gas bridging the cluster-forming gas cores in S235.
In the channel maps (Figure~\ref{VEL}) gas bridges exhibit a typical width of 1.5 times the grid spacing, i.e. $\geq 50 ^{\prime \prime}$ wide (0.4 pc at a distance of 1.56 kpc). The first moment map (Figure~\ref{VVWC}) reveals smooth velocity transitions between the cores indicating a continuous physical link between all 4 cores. The same structures can be seen in the $^{13}$CO and CS channel maps (Figures \ref{13CO} and \ref{CS}) re-imaged from \citet{Kir08}, further supporting this view.

Low excitation ammonia (1,1) is detected in both cores and inter-core bridges, while (2,2) and (3,3) emission is more prominent in the cores. The presence of higher excitation ammonia in the cores indicates that the molecular gas has reached higher column densities and temperatures, suggestive of contraction and internal heating from the resulting star formation. 
The cold gas bridges seen in ammonia show no indication of star formation activity, as is supported by the lack of inter-core stellar density enhancements \citep{Dew11}. 
This is reflected in our integrated spectral analyses in Section~\ref{SPECT} which reveal that the inter-core bridges comprise of gas of lower temperature and column density than core gas.

%Such a cloud would have had a mass roughly equal to the sum of the current cores, i.e. $323 \pm 64$ M$_{\odot}$ using our results or 402 M$_{\odot}$ using the mass estimates of \citet{Kir14}.

Dust filaments bearing some resemblance to those discussed here are seen at smaller scales with a characteristic width of $\sim$ 0.1 pc - in which cores house individual protostars. The configuration, commonly described as `pearls on a string', was initially seen in infrared data from the Herschel space telescope \citep{Az11} and has since been recognised as a common feature of star formation. On the other hand, filaments of much larger scales have been found in other star forming regions such as NGC 6334 \citep{Zernickel13}, and gas bridges of a similar scale to those in S235 connect multiple developed H${\rm II}$ regions in OMC-1 \citep{Hacar17} and the large filament observed in molecular gas in WB673 \citep{Kirsanova17}. However the nature of the aforementioned filamentary systems differs markedly from those seen in S235, which instead connect \emph{clusters} of YSOs rather than individual protostars, and have formed in the presence of-, and by interaction with- a single H${\rm II}$ region. Additionally, analysis of the dense gas physical parameters implies hyper-critical line masses, highlighting the potential for further star formation.

Our interpretation is that the quiescent physical gas bridges linking the cores are the remnants of a large scale fragmentation process in which the cluster-forming cores of S235 condensed out of a single parent molecular cloud. Further evidence of the existence of a natal gas component comes in the way of the absorption feature discussed above. Whether the fragmentation of the parent cloud was driven by CCC, C\&C, or by RDI, should therefore be considered.
\citet{Whitworth94} showed that the swept-up gas layers of expanding nebulae, winds, and CCCs were likely to collapse, by gravitational instability, to form massive cores of gas. Furthermore, \citet{Walch15} showed that C\&C and RDI caused by an expanding H${\rm II}$ region are capable of producing a shell-like structure studded with cores. Their simulations produce a configuration of bridged gas cores similar to those seen in this work.

On the scale of the larger S235ABC complex, \citet{DewanganOjha17} revealed evidence of CCC as a likely trigger of the subsequent star formation seen in this region. \citet{Dew16} showed pressure from the expanding H${\rm II}$ region to be the dominant driver of gas dynamics in the S235 main, capable of explaining the formation of E2, CE and CW (see their Section 3.7), remarking that the youngest core, E1, may be better explained by RDI (a similar conclusion was also reached by \cite{Kir14}).

Considering our result in the context of these works we conclude that the ammonia gas bridges found in S235 likely represent the hyper-critical remnants of CCC-induced fragmentation of a gas cloud involving the C\&C mechanism with likely contribution from the RDI process. Both processes contribute to the proliferation of triggered star formation, driven by the central H${\rm II}$ region of S235.

% \newpage 

\null

\section{Conclusions}

The main conclusions of this paper can be summarised as follows:

\begin{itemize}
\item We performed position-switch mapping observations of the S235 and S235AB regions in ammonia (1,1), (2,2), (3,3) and the 22 GHz water maser transition using the Nobeyama 45m radio telescope.

\item Our observations determined the physical properties of molecular gas in the cores of this SFR, which agree with, and expand on, the previous works in the literature.

\item Focusing on the less-studied gas away from the cores, our observations uncovered the presence of gas bridges that link the cluster-forming cores in the S235 region. These bridges appear to be remnants of a fragmentation event which led to the formation of its present day cores from a larger parent cloud. This fragmentation was likely driven by impact of the extended HII region S235 to surrounding molecular cloud.

\item The presence of dense gas bridges was corroborated by CS and $^{18}$CO gas maps, re-imaged from \citet{Kir08}.

\item Further relic gas was detected in absorption at the foreground of the radio continuum peak in S235 at a velocity consistent with the local diffuse molecular cloud. Thus there are two ammonia gas components in the S235 region: old quiescent gas of low brightness temperature (seen in absorption) and younger, more active star-forming gas which is seen to interact with the H${\rm II}$ region (seen in emission).

\item Our study detected strong water masers associated with star formation in S235AB and the E1 core of S235, the latter being a new maser detection.

\end{itemize}

%%%%%%%%%%%%%%%%%%%%%%%%%%%%%%%%%%%%%%%%%%%%%%%%%%%%%%%%%%
%%%%%%%%%%%%%%%%%%%%%%%%%%%%%%%%%%%%%%%%%%%%%%%%%%%%%%%%%%

\begin{ack}
RB is supported by the East Asia Core Observatory Association (EACOA) under the research fellowship program.
MSK was partly supported by Russian Science Foundation, research project 18-72-10132.
AMS was funded by Russian Foundation for Basic Research through research project 18-02-00917. DAL was supported by the Ministry of Education and Science (the basic part of the State assignment, RK no. AAAA-A17-117030310283-7). This work is partially supported by the Act 211 Government of the Russian Federation, agreement No. 02.A03.21.0006.

\end{ack}

\renewcommand*{\bibfont}{\small}

%%%%%%%%%%%%%%%%%%%%%%%%%%%%%%%%%%%%%%%%%%%%%%%%%%%%%%%%%%
\bibliographystyle{aa}

\end{document}